\documentclass[11pt]{article}

\usepackage{epsfig}
\usepackage{amsmath}
\usepackage{amsfonts}
\usepackage{amssymb}

\newcommand{\proof}{\noindent {\bf Proof:}\ \ }
\newcommand{\qed}{\hfill$\bf{\Box}$}

\newcommand{\IGNORE}[1]{}

\newcommand{\be}{\begin{equation}}
\newcommand{\beq}{\begin{equation}}
\newcommand{\ee}{\end{equation}}
\newcommand{\eeq}{\end{equation}}
\newcommand{\bea}{\begin{eqnarray}}
\newcommand{\eea}{\end{eqnarray}}
\newcommand{\beaa}{\begin{eqnarray*}}
\newcommand{\eeaa}{\end{eqnarray*}}

\newcommand{\bdm}{\begin{displaymath}}
\newcommand{\edm}{\end{displaymath}}

\newtheorem{theorem}{Theorem}
\newtheorem{lemma}[theorem]{Lemma}
\newtheorem{proposition}[theorem]{Proposition}
\newtheorem{corollary}[theorem]{Corollary}

\newtheorem{remark}{Remark}
\newtheorem{assumption}{Assumption}
\newtheorem{algorithm}{Algorithm}
\newtheorem{definition}{Definition}[section]

\newcommand{\dfn}{\stackrel{\triangle}{=}}

\newcommand{\reals}{{\mathbb{R}}}

\newcommand{\argmin}{\mathop{\rm argmin}}
\newcommand{\argmax}{\mathop{\rm argmax}}

\newcommand{\citep}{\cite}
\newcommand{\citet}{\cite}
\newcommand{\DelI}{\Delta(I)}
\newcommand{\DelJ}{\Delta(J)}

\newcommand{\cF}{\mathcal{F}}
\newcommand{\cA}{\mathcal{A}}
\newcommand{\cR}{\mathcal{R}}
\newcommand{\br}{\bar{r}}
\newcommand{\dist}{\mathrm{d}}
\newcommand{\proj}{\mathrm{Proj}}
\newcommand{\W}{W}
\newcommand{\inner}[2]{\langle #1,#2\rangle}
\newcommand{\sign}{{\rm sign}}
\newcommand{\diam}{{\rm diam}}

\begin{document}
\title{An Online Convex Optimization Approach\\ to Blackwell's Approachability}

\author{
Nahum Shimkin \\
Department of Electrical Engineering \\
Technion -- Israel Institute of Technology\\
Haifa 32000, ISRAEL\\
shimkin@ee.technion.ac.il
}


\maketitle

\begin{abstract}

The notion of approachability in repeated games with vector payoffs was introduced by Blackwell in the 1950s, along with geometric conditions for approachability and corresponding strategies that rely on computing {\em steering directions} as projections from the current average payoff vector to the (convex) target set.
Recently, Abernethy, Batlett and Hazan (2011) proposed a class of approachability algorithms that rely on the no-regret properties of Online Linear Programming for
computing a suitable sequence of steering directions. This is first carried out for target sets that are convex cones, and then generalized to any convex set by embedding it in a higher-dimensional convex cone.
In this paper we present a more direct formulation that relies on the support function of the set, along with suitable Online Convex Optimization algorithms, which leads to a general class of approachability algorithms.
We further show that Blackwell's original algorithm and its convergence follow as a special case.
\end{abstract}


\section{Introduction} \label{sec:intro}

Both Blackwell's theory of approachability  and the no-regret framework of online learning address a repeated decision problem in the presence of on an arbitrary (namely, unpredictable) adversary.
The concept of approachability, introduced in \cite{black}, addresses a fundamental
feasibility issue in for repeated matrix games with vector-valued payoffs.
Referring to one player as the {\em agent} and to the other as {\em Nature}, a set $S$ in the payoff space is {\em approachable} by the agent if he can ensure that the average payoff vector converges (with probability 1) to $S$, irrespectively of Nature's strategy.
Blackwell provided in his paper geometric conditions for approachability, which are both necessary and sufficient for {\em convex} target sets $S$, and a corresponding approachability strategy for the agent.
An extensive recent survey of approachability and its implications can be found in   \cite{Perchet2014}, and a textbook exposition is available in \cite{MachlerSolan2013}.

Concurrently, Hannan \cite{hannan} introduced the concept of no-regret play for repeated matrix games. The {\em regret} of the agent is the shortfall of the
cumulative payoff that was actually obtained relative to the one that could have been obtained with the best (fixed) action in hindsight, given Nature's observed actions.
A no-regret strategy, or algorithm, should ensure that the regret
grows sub-linearly in time.
The no-regret criterion has been widely adopted during the last two decades by the machine learning community as a standard measure for the performance of online learning algorithms, and its scope has been greatly extended.
Of specific relevance here is the Online Convex Optimization (OCO) framework, where Nature's discrete action is replaced by the choice of a convex function at each stage, and the agent's decision is a point in a convex set.
The textbook \cite{Cesa2006Book} offers a broad overview of regret and online learning.
Recent surveys of OCO algorithms may be found in \cite{SSS11,HazanSurvey12}.

It is well known that no-regret strategies for repeated games can be obtained
as a special case of the approachability problem.
This was already observed in \cite{black2}; an alternative formulation that
leads to more explicit strategies was proposed in \cite{Hart01}.
More recently, it was shown in \cite{ABH10} that any no-regret algorithm for
the online {\em linear} optimization problem can be used as a basis for an approachability strategy for convex target sets. The online algorithm is used here compute a sequence of {\em steering vectors}, that replace the projection directions used in Blackwell's original algorithm.

The scheme suggested in \cite{ABH10} first considers target sets $S$ that are convex cones. The
generalization to any convex set is carried out by embedding the original target set in a
convex cone in a higher dimensional payoff space.
The present paper proposes a more direct scheme that avoids the above-mentioned embedding.
This is done by invoking the {\em support function} of the target set, along with
well-known relations between this function and the Euclidean distance to the set.
As the support function is convex, the full arsenal of OCO algorithms may be applied to provide the required sequence of steering vectors.

A natural question concerns the relation between Blackwell's original algorithm
and the present framework. We first observe that Blackwell's algorithm is recovered when the standard Follow the Leader (FTL) algorithm is used for the OCO part.
Establishing the (known) convergence of this algorithm via the proposed OCO framework
is a bit more intricate.
First, when the target set has a smooth boundary, we show that FTL guarantees logarithmic rate,
which ''fast" approachability at a rate of $O(\frac{\log T}{T})$.
To address the general case, we further observe that Blackwell's algorithm is still obtained when
a regularized version of FTL is employed, from which the standard
$O(t^{-1/2})$ convergence rate may be deduced.

The paper proceeds as follows.
In Section \ref{sec:2} we recall the relevant background on Blackwell's approachability and Online Convex Optimization.
Section \ref{sec:3} presents the proposed scheme, in the form of a meta-algorithm that relies on a generic OCO algorithm, discusses the relation to the scheme of \cite{ABH10}, and demonstrates a specific algorithm that is obtained by using Generalized Gradient Descent for the OCO algorithm.
In Section \ref{sec:4} we outline the relations with Blackwell's original algorithm, and
provide some concluding remarks.

{\em Notation:} The standard inner product in $\reals^d$ is denoted by $\inner{\cdot}{\cdot}$,
$\|\cdot\|$ is the Euclidean norm, and $\dist(r,S) = \inf_{s\in S} \|r-s\|$ denotes the
corresponding point-to-set distance. Further,
$B_2=\{w\in\reals^d:\|w\|\leq 1\}$ denotes the Euclidean unit ball,
$\Delta(I)$ is the set of probability distributions over a finite set $I$,
$\diam(S)=\sup_{s,s'\in S}\|s-s'\|$ is the diameter of the set $S$,
and $\|\cR-S\| = \sup_{r\in \cR,s\in S}\|r-s\|$ denotes the maximal distance between
points in the sets $\cR$ and $S$.

\section{Model and Background}
\label{sec:2}
We start with a brief of review of Blackwell's approachability and of Online Convex
Programming, focusing on those aspects that are relevant to this paper.

\subsection{Approachability}
Consider a repeated game with \emph{vector-valued} rewards that is played by
two players, the {\em agent} and {\em Nature}.
Let $I$ and $J$ denote the finite action sets of these players, with corresponding
mixed actions $x=(x(1),\dots,x(|I|))\in\DelI$ and $y=(y(1),\dots,y(|J|))\in\DelJ$.
Let $r:I\times J\to \reals^d$ be
the vector-valued reward function of the single-stage game, which is extended to mixed action as
usual through the bilinear function
$$r(x,y)=\sum_{i,j} x(i) y(j) r(i,j)\,.$$
Similarly, we denote
$$ r(x,j)=\sum_{i} x(i) r(i,j) \,.$$

The game is repeated in stages $t=1,2,\dots$, where at stage $t$ actions
$i_t$ and $j_t$ are chosen by the players, and the reward vector $r(i_t,j_t)$ is obtained.
A pure strategy for the agent is a mapping from each possible history
$(i_1,j_1,\dots,i_{t-1},j_{t-1})$ to an action $i_t$, and a mixed strategy is
a probability distribution over the pure strategies. Nature's strategies may be similarly
defined.

As usual, we restrict attention to so-called behavior strategies of the agent,
where the action $i_t$ is drawn randomly according to a mixed action $x_t$,
using independent draws across stages.
Furthermore, to simplify the presentation, we shall state our results and algorithms
in terms of the {\em smoothed} reward vectors $r(x_t,j_t)$, where the reward $r(i,t,j_t)$
is averaged over the mixed action $x_t$.
This will allow us to state the results in simpler sample-path terms,
rather than probabilistic ones;
we further discuss this formulation below after Theorem \ref{thm:1}.

Let
$$ \br_T = \frac{1}{T}\sum_{t=1}^T r(x_t,j_t) $$
denote the $T$-stage average reward vector.
\begin{definition}[Approachability]
\label{def:Approachability}
A closed set $S\subset\reals^d$ is {\bf\em approachable} if there
exists a strategy for the agent and a sequence $\epsilon(T)\to 0$ such that
\be\label{a.s.}
\lim_{T\to\infty} \dist(\br_T,S) \leq  \epsilon(T)
\ee
holds (w.p.\ 1) for any strategy of Nature.
A strategy of the agent that satisfies this property is
an {\em approachability strategy} for $S$.
\end{definition}

\begin{theorem}[Blackwell, 1956]
\label{thm:1}
A closed and convex set $S\subset\reals^d$ is approachable {\em if and only
if} either one of the following equivalent conditions holds:
\begin{itemize}
\item[(i)] For each unit vector $u\in R^d$,
there exists a mixed action $x=x_S(u)\in \DelI$ such that
\be \label{sep1}
\inner{u}{r(x,j)} \leq \sup_{s\in S}\inner{u}{s} \,,
\quad \text{for all\ } j\in J\,.
\ee
\item[(ii)]
For each $y\in\DelJ$ there exists $x\in\DelI$ such that $r(x,y)\in S$.
\end{itemize}
If $S$ is approachable, then the following strategy is an approachability strategy for $S$:\\
For $v\not\in S$, let $u_S(v)$ be the unit vector that points to
$v$ from $\proj_S(v)$, the closet point to $v$ in $S$.
Then, for $t\geq 1$, if $\br_{t}\not\in S$, choose $x_{t+1}=x_S(u_S(\br_{t}))$;
otherwise, choose an arbitrarily action.
\end{theorem}

The approachability strategy introduced by Blackwell has been generalized in
\cite{Hart01}, that essentially allow different norms to be used for the projection unto $S$. Several recent papers have proposed approachability algorithms that depend on Blackwell's dual condition (condition $(ii)$ in the above Theorem) and avoid the projection step altogether (see \cite{BernShim14} and references therein).
The current paper again proposes a generalization of Blackwell's strategy, but from a different viewpoint.

Let us elaborate on the use of the smoothed rewards $r(x_t,j_t)$.
This offers several useful benefits:
\begin{itemize}
\item[1.] As noted, we obtain sample-path bounds rather than probabilistic ones.
\item[2.] We can state results that hold for any sequence $(j_t)$, rather than any
(mixed) strategy of Nature. This is closer to the spirit of Online Algorithms, where
the notion of a randomized choice by Nature may not be meaningful.
\item[3.] As is well known, the difference $\sum_{t=1}^T r(x_t,j_t)- \sum_{t=1}^T r(i_t,j_t)$
is a Martingale difference sequence, hence of order $\sqrt{T}$.
Thus, the difference in the means is of order $\frac{1}{\sqrt{T}}$, and convergence results
derived for the smoothed mean are valid for the non-smoothed one up to that order.
\end{itemize}
We note that the results in \cite{ABH10} are developed for the rewards $r(x_t,y_t)$,
with the mean taken over $y_t$ as well, and the agent is allowed to observe Nature's
mixed action $y_t$ (or at least the mean reward $r(x_t,y_t)$). We avoid making that
extra step and assume that the agent only observes Nature's pure actions $\{j_t\}$.

As the pure actions $i_t$ of the agent do not affect the rewards $r_t=r(x_t,j_t)$,
we may suppress them in the following discussion and focus on the mixed actions $x_t$.
In particular, we restrict attention to strategies of the agent that assign a
mixed action $x_t$ to each sequence $(j_1,\dots,j_{t-1})$ of Nature's actions.
(Note that there is no need to include the past mixed actions $x_1,\dots,x_{t-1}$ in the history sequence, since they may be computed recursively; in practice, however, we will express $x_t$ as a function of past the reward vector sequence $(r(x_k,j_k))_{k< t}$.)
Since there is no randomization involved,
it may be seen that Definition \ref{def:Approachability} is equivalent to the requirement that the bound \eqref{a.s.} holds (deterministically) for any sequence $(j_1,j_2,\dots)$ of Nature's actions.

\subsection{Online Convex Optimization (OCO)}
OCO extends the framework of no-regret learning to function
minimization.
Let $\W$ be a convex and compact set in $\reals^{d}$, and let $\cF$ be a set of convex and uniformly  bounded functions $f:\W\to\reals$.
Consider a sequential decision problem, where at each stage $t\geq 1$ the agent chooses a point
$w_t\in\W$, and then observes a function $f_t\in F$. An {\em Algorithm} for the agent is
a rule for choosing $w_t$, $t\geq 1$, based on the history $\{f_k,w_k\}_{k\leq t-1}$.
The regret of an algorithm $\cA$ is defined as
\be\label{ineq}
{\rm Regret}_T(\cA) = \sup_{f_1,\dots,f_T\in \cF} \left\{\sum_{t=1}^T f_t(w_t) - \min_{w\in\W}\sum_{t=1}^T f_t(w) \right\}\,,
\ee
where the supremum is taken over all possible functions $f_t\in\cF$.
An effective algorithm should guarantee a small regret, and in particular one that grows
sub-linearly in $T$.

The OCO problem  was introduced in this generality in \cite{Zinkevich03}, along
with the following Online Gradient Descent algorithm:
\be\label{OGD}
w_{t+1}=\proj_{\W}( w_t-\eta_t g_t) \,.
\ee
Here $g_t$ is an arbitrary element of $\partial f_t(w_t)$, the
subdifferential of $f_t$ at $w_t$, $(\eta_t)$ is
a diminishing gain sequence, and ${\rm Proj}_{\W}$ denotes the Euclidean projection onto the convex set $\W$.
To state a regret bound for this algorithm, let
$\diam(W)$ denote the diameter of $\W$,
and suppose that all subgradients of the functions $f_t$ are uniformly bounded in norm
by a constant $G$.
\begin{proposition}[Zinkevich, 2003]
\label{thm:OGD}
For the Online Gradient Descent algorithm in \eqref{OGD} with gain sequence
$\eta_t=\frac{\eta}{\sqrt{t}}$, $\eta>0$, the regret is upper bounded by
\be\label{OGDbound}
{\rm Regret}_T({\rm OGD}) \leq (\frac{\diam(W)^2}{\eta}+2\eta G^2)\sqrt{T}.
\ee
\end{proposition}

Several classes of OCO algorithms are now known, as surveyed in \cite{Cesa2006Book,SSS11,HazanSurvey12}. Of particular relevance here is
the Regularized Follow the Leader (RTFL) algorithm, specified by
\be\label{RFTL}
w_{t+1}=\argmin_{w\in\W}
\left(\sum_{k=1}^{t}f_k(w) + R_t(w)    \right)\,,
\ee
where $R_t(w),\,t\geq 1$ is a sequence of regularization functions.
With $R_t\equiv 0$, the algorithm reduces to the basic Follow the Leader (FTL) algorithm,
which does not generally lead to sub-linear regret, unless additional requirements
such as strong convexity are imposed on the functions $f_t$ (we will revisit the
convergence of FTL in Section \ref{sec:4}). For RFTL, we will require the following standard
convergence result.
Recall that a function $R(w)$ over a convex set $\W$ is called {\em $\rho$-strongly convex} if $R(w)-\frac{\rho}{2}\|w\|^2$ is convex there.
\begin{proposition}
\label{prop:RFTL}
Suppose that each function $f_t$ is Lischitz-continuous over $\W$, with Lipschitz coefficient $L_f$.
Let $R_t(w)= \rho_t R(w)$, where $0<\rho_t<\rho_{t+1}$, and the function $R:W\to [0,R_{\max}]$ is Lipschitz continuous with coefficient $L_R$, and is $1$-strongly convex.  Then,
\be\label{RFTLbound}
{\rm Regret}_T({\rm RFTL})
\leq 2 L_f\sum_{t=1}^T \frac{L_f+(\rho_t-\rho_{t-1})L_R}{\rho_{t}+\rho_{t-1}}+ \rho_T R_{\max} \,.
\ee
\end{proposition}
The last bound can be established along the lines of Theorem 2.11 in \cite{SSS11}, which considers
the case of fixed regularization parameters, $\rho_t\equiv \rho_0$.
The proof is outlined in the Appendix.

\section{OCO-Based Approachability}
\label{sec:3}
This section presents the proposed OCO-based approachability algorithm.
We start by introducing the support function and some of its properties,
and expressing Blackwell's separation condition in terms of this function.
We continue to present the proposed meta-algorithm that employs a generic OCO algorithm,
and then provide as an example the specific algorithm that is obtained when
Online Gradient Descent is used as the OCO algorithm.

\subsection{The Support Function}
Let set $S\subset\reals^d$ be a closed and convex set.
The {\em support function} $h_S:\reals^d\to\reals\cup{\{\infty\}}$ of $S$
is defined as
\[
h_S(w) \triangleq \sup_{s \in S} \inner{w}{s}, \quad w \in \reals^d.
\]
It it is evident that $h_S$ is a convex function (as a pointwise supremum over linear functions),
and is positive homogeneous: $h_S(aw)=ah_S(w)$ for $a\geq 0$. Furthermore,
the Euclidean distance from a point $r$ to $S$
can be expressed as
\begin{equation} \label{dist_support}
\dist(r, S) = \max_{w \in B_2} \left\{\inner{w}{r} - h_S(w) \right\},
\end{equation}
where $B_2$ is the closed Euclidean unit ball
(see, e.g., \citet{boyd04}, Section 8.1.3; this
equality may be readily verified using the minimax theorem). It follows that
\be\label{OptProj}
\argmax_{w \in B_2} \left\{\inner{w}{r} - h_S(w) \right\}
=\left\{\begin{array}{ccc}
          0 & : & r\in S \\
          u_S(r) & : & r\not\in S
        \end{array}   \right.
\ee
with $u_S(r)$ as defined in Theorem \ref{thm:1}, namely the unit vector pointing to $r$ from
$\proj_S(r)$.


Blackwell's separation condition in \eqref{sep1} can now be written in terms of the support function, as
$$
\inner{w}{r(x, j)} \leq \sup_{s \in S} \inner{w}{s} \equiv h_S(w) \,.
$$
We thus obtain the following Corollary to Theorem \ref{thm:1}.
\begin{corollary}
A closed and convex set $S$ is approachable {\em if and only if} for every vector $w\in B_2$ there exists
$x\in \DelI$ so that
\begin{equation} \label{SuppFuncCond}
 \inner{w}{r(x,j)} - h_S(w) \leq 0, \quad \forall j\in J.
\end{equation}
\end{corollary}

Note that the last condition can be written as
${\rm val}(w\cdot r)\leq h_S(w)$, where
$$
{\rm val}(w\cdot r) \dfn
\min_{x\in\DelI}\max_{j \in J} \inner{w}{r(x,j)}\,,
$$
the minimax value of the game with the scalar payoff that is obtained by projection the reward vectors $r(i,j)$ onto $w$.
Consequently, a mixed action $x$ that satisfies \eqref{SuppFuncCond}
can be computed as the minimax strategy for the agent in this game.

\subsection{The General Algorithm}

The proposed algorithm builds on the following idea.
First, we employ an OCO algorithm to generate a sequence of {\em steering vectors} $w_t\in B_2$,
so that
\begin{equation}
\label{steering}
\sum_{t = 1}^T \left(\inner{w_t}{r_t} - h_S(w_t) \right)
\geq T \max_{w \in B_2} \left\{\inner{w}{\br_T} - h_S(w) \right\} - a(T),
\end{equation}
where $r_t=r(x_t,j_t)$ is considered an arbitrary vector that is revealed after $w_t$
is specified, and $a(T)=o(T)$. Next, given $w_t$, we choose
$x_t$ that satisfies \eqref{SuppFuncCond}, so that $\inner{w_t}{r_t}-h_S(w_t)\leq 0$.
Using this inequality in \eqref{steering}, and observing the distance formula \eqref{dist_support},
yields
$$
\dist(\br_t,S)\leq \frac{a(T)}{T}\to 0 \,.
$$

To secure \eqref{steering}, observe that the function
$f(w;r) = -\inner{w}{r} + h_S(w)$ is convex in $w$ for each vector $r$.
Therefore, an OCO algorithm can be applied to the sequence of convex functions
$f_t(w) = -\inner{w}{r_t} + h_S(w)$, where $r_t=r(x_t,j_t)$ is considered
an arbitrary vector which is revealed only after $w_t$ is specified.
Applying an OCO algorithm $\cA$  with ${\rm Regret}_T(\cA) \leq a(T)$ to this setup,
we obtain a sequence $(w_t)$ such that
$$
\sum_{t=1}^T f_t(w_t) \leq  \min_{w\in B_2} \sum_{t=1}^T f_t(w) + a(T) \,,
$$
where
\begin{align*}
\sum_{t=1}^T f_t(w_t) &= - \sum_{t=1}^T (\inner{w_t}{r_t} - h_S(w_t)) \,, \\
\sum_{t=1}^T f_t(w) &= -\sum _{t=1}^T (\inner{w}{r_t} - h_S(w)) = - T(\inner{w}{\br_T} - h_S(w))\,.
\end{align*}
This clearly implies \eqref{steering}.

The discussion above leads to the following approachability meta-algorithm.
\begin{algorithm}[Approachability Meta-Algorithm Based on OCO]\ \\
\label{Alg1}
Given:
A closed, convex and approachable set $S$;
a procedure (e.g., a linear program) to compute $x$,  for a given vector $w$,
 so that \eqref{SuppFuncCond} is satisfied;
an OCO algorithm $\cA$ for the functions
$f_t(w)=-\inner{w_t}{r_t}+h_S(w)$, with $\text{Regret}_T(\cA)\leq a(T)$.

Repeat for $t=1,2,\dots$:
\begin{itemize}
\item[1.] Obtain $w_t$ from the OCO algorithm applied to the convex functions $f_k(w)=-\inner{w}{r_k}+h_k(w)$, $k\leq t-1$, so that inequality \eqref{steering} is satisfied.
\item[2.]
Choose $x_t$ according to \eqref{SuppFuncCond}, so that
$\inner{w_t}{r(x_t,j)} - h_S(w_t) \leq 0$ holds for all $j\in J$.
\item[3.]
Observe Nature's action $j_t$, and set $r_t=r(x_t,j_t)$.
\end{itemize}
\end{algorithm}

\begin{proposition}
\label{prop:conv}
For the algorithm above,
$$ \dist(\br_T,S)\leq \frac{a(T)}{T}$$
is satisfied for all $T\geq 1$ and any sequence $(j_1,j_2,\dots)$ of Nature's actions.
\end{proposition}
\proof
As observed above, application of the OCO algorithm implies \eqref{steering}, so that
\begin{align*}
\dist(\br_T,S) &= \max_{w \in B_2} \left\{\inner{w}{\br_T} - h_S(w) \right\} \\
  & \leq  \frac{1}{T} \sum_{t = 1}^T (\inner{w_t}{r_t} - h_S(w_t)) +\frac{a(T)}{T}
  \:\: \leq \:\: \frac{a(T)}{T} \,.
\end{align*}
\qed

To recap, any OCO algorithm that guarantees \eqref{steering} with $\frac{a(T)}{T}\to 0$,
induces an approachability strategy with rate of convergence
$\frac{a(T)}{T}$.

\begin{remark}[Convex Cones]\label{remark:1}
{
The approachability algorithm developed in \cite{ABH10} starts with a target sets $S$ that are
restricted to be convex cones. For $S$ a closed convex cone, the support function is given by
$$
h_S(w)=\left\{\begin{array}{ccc}
                     0 & : & w\in S^o \\
                     \infty & : & w\not\in S^o
                   \end{array}\right.
$$
where $S^o$ is the {\em polar cone} of $S$. The required inequality in \eqref{steering}
therefore reduces to
$$
\sum_{t=1}^T \inner{w_t}{r_t} \geq
T \max_{w\in B_2\cap S^o} \inner{w}{\br_T} - a(T) \,.
$$
The sequence $(w_t)$ can be obtained in this case by applying an online {\em linear} optimization algorithm restricted to $w_t\in B_2\cap S^o$.
This is the algorithm proposed in \cite{ABH10}.

The extension to general convex sets is handled there
by lifting the problem to a $(d+1)$-dimensional space, with payoff vector $r'(x,y)=(\kappa, r(x,y))$ and target set $S'={\rm cone}(\{\kappa\}\times S)$, where $\kappa=\max_{s\in S}\|s\|$, for which it holds that $\dist(u,S)\leq 2 \dist(u',S')$. For further details see \cite{ABH10}.
} 
\end{remark}

\subsection{An OGD-based Approachability Algorithm}

As a concrete example, let us apply the Online Gradient Descent algorithm specified in \eqref{OGD}
to our problem.
With $\W=B_2$ and $f_t(w) = -(\inner{w}{r_t}-h_S(w))$, we obtain
in step 1 of Algorithm \ref{Alg1},
$$
w_{t+1}={\rm Proj}_{B_2}\{w_t + \eta_t(r_t- y_t)\}\,,\quad y_t \in \partial h_S(w_t) \,.
$$
Observe that ${\rm Proj}_{B_2}(v) = v / \max\{1,\|v\|\}$, and
(e.g.,  Corollary 8.25 in \cite{RockWets97})
$$
\partial h_S(w) = \argmax_{s\in S}\inner{s}{w} \,.
$$
To evaluate the convergence rate in \eqref{OGDbound}, observe that $\diam(B_2)=2$,
and, since $y_t\in S$,
$\|g_t\|= \|r_t-y_t\| \leq \|\cR-S\|$, where $\cR=\{r(x,y)\}_{x\in\DelI,y\in\DelJ}$
is the reward set. Assuming for the moment that the goal set $S$ is bounded, we obtain
$$
\dist(\br_T,S) \leq \frac{b(\eta)}{\sqrt{T}}\,, \quad {\rm with\ \ } b(\eta) = \frac{4}{\eta}+2\eta \|\cR-S\|^2 \,.
$$
For $\eta=\sqrt{2}/\|\cR-S\|$, we thus obtain $b(\eta)=4\sqrt{2}\|\cR-S\|$.

If $S$ is not bounded, it can always be intersected with $\cR$ (without affecting its approachability), yielding $\|\cR-S\|\leq \diam(\cR)$.
This amounts to modifying the choice of $y_t$ in the algorithm to
$$
y_t\in \partial h_{S\cap\cR}(w_t) = \argmax_{y\in S\cap\cR}(y,w) \,.
$$
Alternatively, one may restrict attention
(by projection) to vectors $w_t$ in the set $\{w\in B_2 : h_S(w)<\infty\}$, similarly to the case of convex cones mentioned in Remark \ref{remark:1} above; we will not go into further details here.

\section{Blackwell's Algorithm and (R)FTL}
\label{sec:4}
We next examine the relation between Blackwell's approachability algorithm and the present OCO-based framework. We first show that Blackwell's algorithm coincides with OCO-based approachability
when FTL is used as the OCO algorithm. We use this equivalence to establish fast (logarithmic) convergence rates for Blackwell's algorithm when the target set $S$ has a smooth boundary.
Interestingly, this equivalence does not provide a convergence result for general convex sets.
To complete the picture, we show that Blackwell's algorithm can more generally be obtained via a {\em regularized} version of FTL, which leads to an alternative proof of convergence of the algorithm in the general case.

\subsection{Blackwell's algorithm as FTL}
Recall Blackwell's algorithm as specified in
Theorem \ref{thm:1}, namely $x_{t+1}$ is chosen as a mixed action that satisfies \eqref{sep1} for $u=u_S(\br_{t})$.
\begin{lemma} \label{lemma:equiv1}
For $f_t(w)=-\inner{w}{r_t}+h_S(w) $,
$$
\argmin_{w\in B_2}\sum_{k=1}^t f_k(w)=
\left\{\begin{array}{lcl}
    u_S(\br_{t})  & : & \br_t\not\in S \\
    0 & : & \br_t\in S
\end{array} \right. \,.
$$
\end{lemma}
\proof
Observe that $\sum_{k=1}^t f_k(w) = -t(\inner{w}{\br_t}-h_S(w))$, so that
$$  \argmin_{w\in B_2}\sum_{k=1}^t f_k(w) = \argmax_{w\in B_2}\{\inner{w}{\br_t}-h_S(w) \} \,. $$
The required equality now follows from \eqref{OptProj}.
\qed

Comparing to \eqref{RFTL}, with $R_t\equiv  0$, it may be seen that the sequence of projection directions $u_S(\br_{t})$ in Blackwell's algorithm coincides with the
sequence $(w_t)$ that is obtained by applying the FTL algorithm to the functions $(f_t)$ over $w\in B_2$. It follows that Blackwell's algorithm is identical to Algorithm \ref{Alg1} with this choice of the OCO algorithm.

To establish convergence of Blackwell's algorithm via this equivalence, one needs to show that FTL guarantees the regret bound in \eqref{steering} for an arbitrary reward sequence
$(r_t)\subset\cR$, with a sublinear rate sequence $a(T)$.
It is well know, however, that (unregularized) FTL does not guarantee sublinear regret,
without some additional assumptions on the function $f_t$.
A simple counter-example, reformulated to the present case, is devised as follows:
Let  $S=\{0\}\subset \reals$, so that $h_S(w)=0$, and suppose that
$r_1=-1$ and $r_t=2(-1)^t$ for $t>1$. Since $w_t = \sign(\br_{t-1})$ and
$\sign(r_t)=-\sign(\br_{t-1})$, we obtain
that $f_t(w_t)=-r_tw_t=1$, leading to a linearly-increasing regret.

The failure of FTL in this example is clearly due to the fast changes in the
predictors $w_t$.
We now add some smoothness assumptions on the set $S$ that can mitigate such abrupt changes.
\begin{assumption}
\label{assumption:smooth}
Let $S$ be a compact and convex set. Suppose that the boundary $\partial S$ of $S$ is
smooth with curvature bounded by $\kappa_0$, namely:
\be\label{curvature}
\|\vec{n}(s_1)-\vec{n}(s_2)\| \leq \kappa_0 \| s_1-s_2\|\quad \text{for all\ \ } s_1,s_2\in\partial S \,,
\ee
where $\vec{n}(s)$ is the unique unit outer normal to $S$ at $s\in \partial S$.
\end{assumption}
For example, for a closed Euclidean ball of radius $\rho$,
\eqref{curvature} is satisfied with equality for $\kappa_0=\rho^{-1}$.
The assumed smoothness property may in fact be formulated in terms of an interior sphere condition:
For any point in $s\in S$ there exists a ball $B(\rho)\subset S$ with radius $\rho=\kappa_0^{-1}$ such that $s\in B(\rho)$.

\begin{proposition}
\label{prop:log}
Let Assumption \ref{assumption:smooth} hold. Consider Blackwell's algorithm as specified in Theorem \ref{thm:1},
and denote $w_t = u_S(\bar{r}_{t-1})$ (with $w_1$ arbitrary).
Then, for any time $T\geq 1$ such that $\br_T\not\in S$, \eqref{steering} holds with
\be\label{log_bound}
a(T) =C_0 (1+\ln T),
\ee
where $C_0= \diam(\cR)\, \|\cR-S\| \, \kappa_0$, $C_1=\|\cR-S\|$, and $\ln(\cdot)$ is the natural
logarithm.
Consequently,
\be\label{convFTL}
\dist(\br_T,S) \leq C_0 \frac{1+\ln T}{T} \,, \quad T\geq 1\,.
\ee
\end{proposition}
\proof See the Appendix.

The last result establishes a fast  convergence rate (of order $\log T/T$) for Blackwell's
approachability algorithm, under the assumed smoothness of the target set.
We observe that in the stochastic version of the algorithm, which is based on the rewards $r(i_t,j_t)$ rather than $r(x_t,j_t)$, the convergence is still of order
$T^{-1/2}$ due to the added stochastic effect (unless all mixed actions $x_t$ happen to be pure).
We also note that logarithmic convergence rates for OCO algorithms were derived in
\cite{log_reg07}, under strong convexity conditions on the function $f_t$.
Finally, conditions for fast approachability
(of order $T^{-1}$) were derived in \cite{PerchetMannor13}, but are of different nature than the above.

\subsection{Blackwell's algorithm as RFTL}

The smoothness requirement in Assumption \ref{assumption:smooth} precludes such important
target sets as polyhedra and cones.
As observed above, in absence of such additional smoothness properties
the interpretation of Blackwell's algorithm through an FTL scheme
does not imply its convergence, as the regret of FTL (and the corresponding bound $a(T)$ in \eqref{steering}) might increase linearly in general.

To address the general case, we show next that the Blackwell's algorithm can be identified
more generally with a {\em regularized} version of FTL.
This algorithm does guarantee an $O(\sqrt{T})$ regret in \eqref{steering}, and consequently leads to
the standard $O(T^{-1/2})$ rate of convergence of Blackwell's approachability algorithm.

Our starting point is the following observation:
\begin{lemma} \label{lemma:equivalence2}
For $f_k(w)=-\inner{w}{r_k}+h_S(w)$, $1\leq k \leq t$, and any $\rho_t>0$,
\be\label{opt2}
w_{t+1} \dfn \argmin_{w\in B_2}\big\{ \sum_{k=1}^t f_k(w) +\frac{\rho_t}{2}\|w\|^2 \big\}=
\left\{\begin{array}{lcl}
    \beta_t u_S(\br_{t})  & : & \br_t\not\in S \\
    0 & : & \br_t\in S
    \end{array} \right. \,.
\ee
where  $\beta_t = \min\{1,\frac{t}{\rho_t}\dist(\br_t,S)\} >0$.
\end{lemma}
\proof
Recall that $\sum_{k=1}^t f_k(w) = -t(\inner{w}{\br_t}-h_S(w))$, so that
$$
\argmin_{w\in B_2}\big\{ \sum_{k=1}^t f_k(w) +\frac{\rho_t}{2}\|w\|^2 \big\} = \argmax_{w\in B_2}\{\inner{w}{\br_t}-h_S(w) -\frac{\rho_t}{2t} \|w\|^2 \} \,.
$$
To compute the right-hand side, we first maximize over $\{w:\|w\|=\beta\}$, and then
optimize over $\beta\in [0,1]$. Denote $r=\bar{r}_t$, and $\eta=\rho_t/t$.
Similarly to Lemma \ref{lemma:equiv1},
$$
\argmax_{\|w\|=\beta}\{\inner{w}{r}-h_S(w) -\frac{\eta}{2} \|w\|^2 \}
=\argmax_{\|w\|=\beta}\{\inner{w}{r}-h_S(w)\}
=\left\{\begin{array}{lcl}
    \beta u_S(r)  & : & r\not\in S \\
    0 & : & r\in S
    \end{array} \right. \,.
$$
Now, for $r\not\in S$,
$$
\max_{\|w\|=\beta}\{\inner{w}{r}-h_S(w) -\frac{\eta}{2} \|w\|^2 \} =
\beta\dist(r,S)- \frac{\eta}{2}\beta^2 \,.
$$
Maximizing the latter over $0\leq \beta \leq 1$ gives
$\beta^* = \min\{1,\frac{\dist(r,S)}{\eta}\}$.
Substituting back $r$ and $\eta$ gives \eqref{opt2}.
\qed

Equation \eqref{opt2} defines an RFTL algorithm with quadratic regularization.
When used for the OCO part in Algorithm \ref{Alg1}, the resulting scheme turns out to
be equivalent to Blackwell's algorithm.
Indeed, the minimum in \eqref{opt2} is attained by the same unit vector
$u_S(\br_t)$ that appears in Theorem \ref{thm:1}, scaled by a positive
constant. That scaling does not affect the choice of $x_t$ according to
\eqref{SuppFuncCond}, as the support function $h_S(w)$ is positive homogeneous.
However, this scaling does induce sublinear-regret for the OLO algorithm, and consequently convergence of the approachability algorithm. This is summarized as
follows.
\begin{proposition}
\label{prop:sqrt}
Let $S$ be a convex and compact set. Consider the RTFL algorithm specified in equation \eqref{opt2}, with $\rho_t = \rho\sqrt{t}$, $\rho>0$. The regret of this algorithm is bounded by
$$
\mathrm{Regret}_T(\mathrm{RTFL}) \leq
(\frac{2L_f^2}{\rho}+\rho)\sqrt{T}  + \frac{2L_f^2}{\rho}+ L_f\ln(4T-3)
\;\; \dfn\;\; a_0(T)   \,,
$$
where $L_f=\|\cR-S\|$.
Consequently, if this RTFL algorithm is used in step 1 of Algorithm \ref{Alg1} to
provide $w_t$, we obtain
\be\label{convRFTL}
\dist(\br_T,S) \leq \frac{a_0(T)}{T} = O(T^{-\frac{1}{2}}) \,, \qquad T\geq 1\,.
\ee
\end{proposition}
\proof
The regret bound follows from the one in Proposition \ref{prop:RFTL},
evaluated for $f_t(w)= -\inner{r_t}{w}+h_S(s)$, $W=B_2$, $R(w)=\|w\|^2$, and $\rho_t=\rho_0\sqrt{t}$.  Recalling that
 $\partial f_t(w) = -r_t+\argmax_{s\in S}\inner{w}{s}$,
 the Lipschitz constant of $f_t$ is upper bounded by $\|\cR-S\|\dfn L_f$.
Furthermore, $R_{\max}=1$ and $L_R=2$. Therefore,
$$
\mathrm{Regret}_T(\mathrm{RTFL})
\leq 2 L_f\sum_{t=1}^T
\frac{L_f+2\rho(\sqrt{t}-\sqrt{t-1})}{\rho(\sqrt{t}+\sqrt{t-1})} +\rho\sqrt{T} \,.
$$
Upper bounding the sums with corresponding integrals gives the stated regret bound.
The second part now follows directly from Proposition \ref{prop:conv}.
\qed

With $\rho=\sqrt{2}L_f$, we obtain in \eqref{convRFTL} the convergence rate
$$
\dist(\br_T,S) \leq \frac{2\sqrt{2}\|\cR-S\|}{\sqrt{T}}+o(\frac{1}{\sqrt{T}}) \,.
$$

We emphasize that the algorithm discussed in this section is equivalent to Blackwell's
algorithm, hence its convergence is well known.
The proof of convergence here is certainly not the simplest, nor does it lead to the best constants in the convergence rate.
Indeed, Blackwell's proof (which recursively bounds the square distance $\dist(\br_T,S)^2$)
leads to the bound $\dist(\br_T,S) \leq \frac{\|\cR-S\|}{\sqrt{T}}$.
Rather, our main purpose here was to provide an alternative view and analysis of Blackwell's algorithm, which rely on a standard OCO algorithm.
Nonetheless, the logarithmic convergence rate that was obtained under the smoothness
Assumption \ref{assumption:smooth} appears to be new.

\section*{Acknowledgements}
The author wishes to thank Elad Hazan for helpful comments
on a preliminary version of this work.
This research was supported by the Israel Science
Foundation grant No.\ 1319/11.

\bibliography{letter_bib}


\appendix
\section*{Appendix}

{\bf Proof of Proposition \ref{prop:RFTL}:\ \ }
We follow the outline of the proof of Lemma 2.10 in \cite{SSS11}, modified to accommodate
a non-constant regularization sequence $\rho_t$. The starting point is the inequality, proved by
induction,
\be\label{induct}
\sum_{t=1}^T(f_t(w_t)-f_t(u)) \leq \sum_{t=1}^T(f_t(w_t)-f_t(w_{t+1})) +\rho_t R(u) \,,
\ee
which holds for any $u\in \W$. Therefore,
\be\label{bound1}
\sum_{t=1}^T(f_t(w_t)-f_t(u)) \leq L_f \sum_{t=1}^T\|w_t-w_{t+1} \| +\rho_t R(u) \,.
\ee

Denote $F_t(w) = \sum_{k=1}^{t-1}f_k(w) + \rho_{t-1}R(w)$. Then $F_t$ is $\rho_{t-1}$-strongly convex, and $w_{t}$ is its maximizer by definition. Hence, it holds generally that
$$
F_t(u) \geq F_t(w_t) +\frac{\rho_{t-1}}{2}\|u-w_t\|^2 \,,
$$
and in particular,
\begin{align}
F_t(w_{t+1}) & \geq F_t(w_t) +\frac{\rho_{t-1}}{2}\|w_{t+1}-w_t\|^2,\\
F_{t+1}(w_{t}) & \geq F_{t+1}(w_{t+1}) +\frac{\rho_{t}}{2}\|w_{t}-w_{t+1}\|^2 \,.
\end{align}
Summing and cancelling terms, we obtain
$$
f_t(w_t) - f_t(w_{t+1}) +(\rho_t-\rho_{t-1})(R(w_t)-R(w_{t+1}) \geq \frac{\rho_{t}+\rho_{t-1}}{2} \|w_{t+1}-w_t\|^2 \,.
$$
But the left-hand side is upper-bounded by $(L_f+(\rho_t-\rho_{t-1})L_R)\|w_{t+1}-w_t \|$,
which implies that
$$
\|w_{t+1}-w_t\| \leq 2\frac{L_f+(\rho_t-\rho_{t-1})L_R}{\rho_{t}+\rho_{t-1}} \,.
$$
Substituting in \eqref{bound1} gives the bound stated in the Proposition.
\qed

{\bf Proof of Proposition \ref{prop:log}:\ \ }
We first observe that the regret bound in \eqref{log_bound} implies \eqref{convFTL}.
Indeed, for $\br_T\not\in S$,  $\dist(\br_T,S)\leq a(T)/T$
follows as in Proposition \ref{prop:conv}, while if $\br_T\in S$ then
$\dist(\br_T,S)=0$ and \eqref{convFTL} holds trivially.

We proceed to establish the logarithmic regret bound in \eqref{log_bound}.
Let $f_t(w) = -\inner{w}{r_t} +h_S(w)$, $W=B_2$,  and denote
\be\label{prf1}
{\rm Regret}_T(f_{1:T}) =
 \sum_{t=1}^T f_t(w_t) - \min_{w\in\W}\sum_{t=1}^T f_t(w) =
 \sum_{t=1}^T(f_t(w_t)-f_t(w_{T+1})) \,.
 \ee
A standard induction argument (e.g., Lemma 2.1 in \cite{SSS11}) verifies that
 \be\label{prf2}
\sum_{t=1}^T(f_t(w_t)-f_t(u)) \leq
 \sum_{t=1}^T(f_t(w_t)-f_t(w_{t+1}))
\ee
holds for any $u\in\W$, and in particular for $u=w_{T+1}$. It
remains to upper-bound the differences in the last sum.

Consider first the case where $\br_t\not\in S$ for all $1\leq t\leq T$. We first show that
$\|w_t-w_{t+1}\|$ is small, which implies the same for $|f_t(w_t)-f_t(w_{t+1})|$.
By its definition, $w_{t+1}=u_S(\br_t)$, the unit vector pointing to $\br_{t}$ from
$c_t\dfn\proj_S(\br_t)$, which clearly coincides with the outer unit normal $\vec{n}(c_{t})$
to S at $c_t$.
It follows that
$$
\|w_t-w_{t+1}\| =\| \vec{n}(c_{t-1}) - \vec{n}(c_{t}) \| \leq \kappa_0 \|c_{t-1}-c_{t}\|
\leq \kappa_0 \|\br_{t-1}-\br_{t}\| \,,
$$
where the first inequality follows by Assumption \ref{assumption:smooth}, and the second
due to the shrinking property of the projection. Substituting
$\br_{t}=\br_{t-1}+\frac{1}{t}(r_t-\br_{t-1})$ obtains
\be\label{ineq1}
\|w_t-w_{t+1}\| \leq  \frac{\kappa_0}{t} \| r_t-\br_{t-1} \| \leq \frac{\kappa_0}{t} \diam(\cR) \,.
\ee
Next, observe that for any pair of unit vectors $w_1$ and $w_2$,
\begin{align*}
f_t(w_1)-f_t(w_2) &= -\inner{w_1-w_2}{r_t} + h_S(w_1) - h_S(w_2) \\
 &= -\inner{w_1-w_2}{r_t} + \max_{s\in S} \inner{w_1}{s} - \max_{s\in S} \inner{w_2}{s} \\
 &\leq  -\inner{w_1-w_2}{r_t} +  \inner{w_1}{s_1} -  \inner{w_2}{s_1} \\
 &= \inner{w_1-w_2}{s_1 -r_t} \;\leq\;  \|w_1-w_2\| \|\cR-S\| \,,
\end{align*}
where $s_1\in S$ attains the first maximum. Since the same bound holds for
$f_t(w_2)-f_t(w_1)$, it holds also for the absolute value.
In particular,
\be\label{ineq2}
| f_t(w_t)-f_t(w_{t+1}) | \leq  \|w_t-w_{t+1}\| \|\cR-S\| \,,
\ee
and together with \eqref{ineq1} we obtain
$$
| f_t(w_t)-f_t(w_{t+1}) | \leq \frac{\kappa_0}{t} \, \diam(\cR)\,\|\cR-S\| = \frac{C_0}{t} \,.
$$
Substituting in \eqref{prf2} and summing over $t^{-1}$ yields the regret bound
\be\label{no_brt}
{\rm Regret}_T(f_{1:T})\leq C_0(1+\ln T).
\ee
We next extend this bound to case where $\br_t\in S$ for some $t$. In that case
$w_{t+1}=0$, and $w_t-w_{t+1}$ may not be small. However, since $f_t(0)=0$, such
terms will not affect the sum in \eqref{prf2}. Recall that we need to establish
\eqref{log_bound} for $T$ such that $br_T\not\in S$. In that case,
any time $t$ for which $\br_t\in S$ is follows by some
time $m\leq T$ with $\br_m \not\in S$.
Let
$1\leq k < m \leq T$ be indices such that
$\br_k,\dots\br_{m-1}\in S$, but $\br_{k-1}\not\in S$ (or $k=1$) and $\br_m \not\in S$. Then $w_{k+1},\dots,w_m=0$, and
$$
\sum_{t=k}^m (f_t(w_t) - f_t(w_{t+1}) = f_k(w_k) - f_{m}(w_{m+1}) \,.
$$
Proceeding as above, we obtain similarly to \eqref{ineq1},
$$
\|w_k-w_{m+1}\| \leq \kappa_0 \| \br_{k-1} - \br_m \|
\leq \diam(\cR) \sum_{t=k}^{m-1} \frac{\kappa_0}{t} \,,
$$
and the regret bound in \eqref{no_brt} may be obtained as above.
\qed

\end{document}